\documentclass[a4paper]{aaold}
\usepackage{amssymb}
\usepackage{graphicx}

\begin{document}

\title{Stability of magnetic configurations containing the toroidal and  
       axial fields} 
\author{A.~Bonanno\inst{1,2}, V.~Urpin\inst{1,3}}
\offprints{A.Bonanno}
\institute{$^{1)}$ INAF, Osservatorio Astrofisico di Catania,
           Via S.Sofia 78, 95123 Catania, Italy \\
           $^{2)}$ INFN, Sezione di Catania, Via S.Sofia 72,
           95123 Catania, Italy \\
           $^{3)}$ A.F.Ioffe Institute of Physics and Technology and
           Isaac Newton Institute of Chile, Branch in St. Petersburg,
           194021 St. Petersburg, Russia}

\date{\today}

\abstract{Stability properties of  magnetic-field configurations containing the 
toroidal and axial field are considered.  
The stability is treated by making use of  linear analysis. 
It is shown that the conditions required 
for the onset of instability are essentially different from those 
given by the necessary condition $d (s B_{\varphi})/ds > 0$, where $s$ is the 
cylindrical radius. The growth rate of instability is calculated for a wide 
range of the parameters. We argue that the instability can operate in two 
different regimes depending on the strength of the axial field and the 
profile of the toroidal field.
}
\authorrunning{A.Bonanno \& V.Urpin}
\titlerunning{Stability of the magnetic field}

\maketitle

\section{Introduction}

Turbulence generated by MHD instabilities can play an important role in 
enhancing transport processes in various astrophysical bodies, such as 
accretion and protoplanetary disks, galaxies, stellar radiative zones, etc.  
The anomalous turbulent transport can be particularly important in magnetized 
gas where a wide variety of MHD instabilities  can occur (see, e.g., Barnes 
et al. 1999). In this case, the onset of instability can be caused both by 
hydrodynamic motions (for instance, differential rotation; see, e.g., 
Velikhov 1959; Chandrasekhar 1960) or unstable magnetic configurations. { 
Which field strength and topology can sustain a stable magnetic configuration 
is still rather uncertain despite extensive work (see Borra et al. 1982; 
Mestel 1999 for review).} 

Most likely, the best-studied magnetic configuration is one with a purely 
toroidal field. Ever since the paper by Tayler (1973), it has been known that toroidal 
fields can be unstable close to the axis of symmetry, if there is a non-zero 
electric current density on the axis. The growth rate of this instability is 
expected to be of the order of the time taken for an Alfv\'en wave to travel 
around the star on a toroidal field line. However, even a purely toroidal 
field is stable if it decreases rapidly with the cylindrical radius $s$. For 
instance, Tayler (1973; see also Chanmugam 1979) argued that the toroidal 
field $B_{\varphi}$ is stable against axisymmetric perturbations if it satisfies 
the condition $d ( B_{\varphi}/s)/ds < 0$ and to non-axisymmetric 
perturbations if $d (s B_{\varphi}^2)/ds <0$. Note that a purely toroidal 
field can also be subject to the magnetic buoyancy instability (Parker 1955; 
Gilman 1970; Acheson 1978) but the Tayler instability likely appears first 
as the strength of the toroidal field increases (Spruit 1999).  
  
The stability of a purely toroidal field in the radiative zones of stars and
accretion disks has been studied by a number of authors. Numerical modeling 
by Braithwaite (2006) confirms that the toroidal field with $B_{\varphi} 
\propto s$ or $\propto s^2$ is unstable to the $m=1$ mode ($m$ is the 
azimuthal wave number) as predicted by Tayler (1973). The linear 
stability of the toroidal field in rotating stellar interiors has been
considered by  Kitchatinov \& R\"udiger (2007), who conclude that the
magnetic instability is essentially three-dimensional and that the finite 
thermal conductivity has a strong destabilizing effect. Terquem \& 
Papaloizou (1996) and Papaloizou \& Terquem (1997) considered the stability 
of an accretion disk with an embedded toroidal magnetic field. These
authors find that the disks containing a purely toroidal field are always 
unstable and obtained spectra of unstable modes in the local approximation.
They argue that one type of modes is driven primarily by buoyancy, while
the other is driven by shear independently of the magnetic configuration. 

{ Stability properties of purely poloidal magnetic configurations  have
also been well-studied. Since the papers by Wright (1973) and Markey \& Tayler 
(1973, 1974), it is known that a poloidal field is subject to dynamical 
instabilities in the neighbourhood of neutral points/lines if the field 
lines are closed inside the star. These authors recognizes that the magnetic
field in the neighbourhood of a neutral line resembles that of a toroidal
pinched discharge that is known to be unstable. Although instabilities
involving significant displacements in the direction of gravity were strongy
inhibited, other instabilities were not affected. The instability of
poloidal configurations is rather fast: its growth time can reach a few
Alfv\'en crossing time (Van Assche et al. 1982; Braithwaite \& Spruit 2006). 
With numerical simulations, the stability of poloidal magnetic configurations
has been studied by Braithwaite \& Spruit (2006), who apply the results to the 
internal magnetic configuration of neutron stars. Note, however, that a 
toroidal field might exert a stabilising influence on the instabilities of
a poloidal field in the neighbourhood of neutral points (Tayler 1980).}

{ On the contrary, the addition of even a relatively weak poloidal field 
alters the stability of the toroidal field substantially. For example, as  
first shown by Howard \& Gupta (1962; see also Knobloch 1992; Dubrulle 
\& Knobloch 1993), a necessary (but not sufficient) condition for the 
instability of a toroidal field in the presence of the axial field reads} 
\begin{equation}\label{1}
\frac{d}{ds} (s B_{\varphi}) > 0.
\end{equation}    
Howard \& Gupta (1962) argue that, for a fixed value of $m$, the growth rate 
of instability caused by condition (1) must vanish in the limit of a 
vanishing axial magnetic field, thereby providing a connection with the 
stability criterion obtained by Tayler (1973). Note that the presence of
a radial field is also crucial for stability properties of rotating magnetic
configurations (Bonanno \& Urpin 2006). { It turns out that configurations 
containing both toroidal and poloidal fields are more stable than purely 
toroidal or purely poloidal ones (Prendergast 1956; Tayler 1980). With 
numerical simulations, Braithwaite \& Nordlund (2006) studied the stability of 
a random initial field in the stellar radiative zone. The star was modeled 
on a Cartesian grid, and the authors found that the stable magnetic 
configurations generally have the form of tori with comparable poloidal and 
toroidal field strengths.} 

In the present paper, we address the stability properties of magnetic 
configurations by considering the stability of the toroidal magnetic field 
with respect to axisymmetric perturbations in the presence of axial fields of 
a various strength. We show that the instability may occur in such magnetic 
field configurations under the conditions that differ substantially from 
those imposed by the Tayler criterion or the necessary condition (\ref{1}). 
We argue that, in some cases, the instability is caused by the new type of 
MHD waves with the growth rate proportional to $\sqrt{B_z B_{\varphi}}$ where 
$B_z$ is the axial magnetic field. Depending on the profile $B_{\varphi}(s)$
and the ratio $B_z/B_{\varphi}$, the instability can occur in two regimes
that have substantially different growth rates. We also show that the range 
of unstable wavelength in the $z$-direction can be essentially different, 
depending on the $B_z/B_{\varphi}$ ratio.       

\section{Basic equations}

{ Let us consider the stability of an axisymmetric cylindrical magnetic
configuration assuming plasma to be conducting perfectly. We work in
cylindrical coordinates ($s$, $\varphi$, $z$) with the unit vectors 
($\vec{e}_{s}$, $\vec{e}_{\varphi}$, $\vec{e}_{z}$). The inner and outer
radii of the configuration are $s_1$ and $s_2$, respectively ($s_1$ can be 
equal to 0). We assume that the axial magnetic field $B_z$ and the azimuthal 
field $B_{\varphi}$ depend on the cylindrical radius alone: $B_z=B_z(s)$ and 
$B_{\varphi}= B_{\varphi}(s)$. Usually, the axial field in our model can 
even change the sign at some point $s_0$ within ($s_1$, $s_2$), $s_2 >s_0 >
s_1$. Such a dependence is considered in order to mimic the stability 
properties of stars where the axial field component can generally change 
the sign.}  

The equations of incompressible MHD are  
\begin{eqnarray}
\frac{\partial \vec{v}}{\partial t} + (\vec{v} \cdot \nabla) \vec{v} = 
- \frac{\nabla p}{\rho} 
+ \frac{1}{4 \pi \rho} (\nabla \times \vec{B}) \times \vec{B}, 
\end{eqnarray}
\begin{equation}
\nabla \cdot \vec{v} = 0, 
\end{equation}
\begin{equation}
\frac{\partial \vec{B}}{\partial t} - \nabla \times (\vec{v} \times \vec{B}) 
= 0,
\end{equation}
\begin{equation}
\nabla \cdot \vec{B} = 0. 
\end{equation}
It is assumed that gas is in hydrostatic equilibrium in the basic state and  that
there is no rotation, so
\begin{equation}
\frac{\nabla p}{\rho} = \frac{1}{4 \pi \rho} 
(\nabla \times \vec{B}) \times \vec{B}.
\end{equation}
In this paper we consider the stability of axisymmetric perturbations. Since
the basic state is stationary, the time dependence of perturbations can be
taken in the form $\exp{\sigma t}$. Small perturbations will be indicated by 
subscript 1, while unperturbed quantities will have no subscript. The 
linearized MHD equations are
\begin{equation}
\sigma \vec{v}_{1} = - \frac{\nabla p_{1}}{\rho} + 
\frac{1}{4 \pi \rho}[ (\nabla \times \vec{B}_{1}) \times \vec{B} 
+ (\nabla \times \vec{B}) \times \vec{B}_{1}], 
\end{equation}
\begin{equation}
\nabla \cdot \vec{v}_{1} = 0, 
\end{equation}
\begin{equation}
\sigma \vec{B}_{1} - (\vec{B} \cdot \nabla) \vec{v}_{1} + (\vec{v}_{1}
\cdot \nabla) \vec{B} = 0,
\end{equation}
\begin{equation}
\nabla \cdot \vec{B}_{1} = 0. 
\end{equation}
{ Equations.~(7)-(10) are homogeneous in $z$ and admit solutions in the form of 
waves in the $z$-direction, $\propto \exp( -ik z)$ where $k$ is the 
wavevector. In stellar conditions, this analysis applies if $k$ satisfies
the condition $k H > 1$ where $H$ is the lengthscale of the basic state in 
the $z$-direction. Eliminating all variables in favour of the radial velocity 
perturbation $v_{1s}$, we obtain the following differential equation
\begin{eqnarray}
(\sigma^2 + \omega_A^2) \left[ \! \frac{d}{d s} \! \left( \frac{d v_{1s}}{ds} 
+ \frac{v_{1s}}{s} \! \right) \! - \! k^2 v_{1s} \right] +
\frac{4 k^2 \omega_{A}^2 \omega_{B}^2}{(\sigma^2 + \omega_{A}^2)} v_{1s} 
\nonumber \\
- 2 k^2 \omega_{B}^2  ( 1 - \alpha )
v_{1s} = - \frac{2}{s} \delta \omega_{A}^2 \left( 
\frac{\partial v_{1s}}{\partial s} + \frac{v_{1s}}{s} \right),
\end{eqnarray} 
where
\begin{equation}
\omega_{A} \! = \! \frac{k B_z}{\sqrt{4 \pi \rho}} , \;\;
\omega_{B} \! = \! \frac{B_{\varphi}}{s \sqrt{4 \pi \rho}} , \;\;
\alpha \! = \! \frac{\partial \ln B_{\varphi}}{\partial \ln s} , \;\;
\delta \! = \! \frac{\partial \ln B_{z}}{\partial \ln s}.
\end{equation}
In the case $\delta=0$ ($B_z=$const), Eq.~(11) recovers the equation derived 
by Acheson (1973) and Knobloch (1992). Once  the radial velocity is known, one 
can calculate the perturbations of other quantities. For example, a 
perturbation of the vertical field can be expressed in terms of $v_{1s}$ as 
$$
B_{1z}= - (B_z/ \sigma s) \partial (s v_{1s})/\partial s - v_{1s} 
\frac{\partial B_{z}}{\partial s}.
$$
With appropriate boundary conditions, Eq.~(11) allows the 
eigenvalues $\sigma$ to be determined. In this study, we choose the simplest boundary 
conditions and assume that the radial velocity vanishes at $s_1$ and $s_2$.}

\section{Analytical consideration of Eq.~(11)}

{ In this section, we consider few particular cases of Eq.~(11) when it 
allows an analytical solution. The analytical consideration of simple cases 
can be a useful guide in understanding the stability properties of more 
complex field configurations as well as in the interpretation of numerical 
results.}  

\subsection{Instability in the case $B_{\varphi} \propto s$.}

{ Let us initially consider the case of the azimuthal field being 
proportional to the cylindrical radius, $B_{\varphi} = B_{\varphi 0} 
(s/s_1)$,  where $B_{\varphi 0}$ is the field strength at $s=s_1$.} Then,
the quantities $B_{\varphi}/s$ and $\omega_{B}$ are constant, and the last 
term on the left hand side of Eq.~(11) vanishes. For the sake of simplicity, we 
also assume  that the axial field is constant. After some algebra, Eq.~(11) 
can be transformed into
\begin{equation}
\frac{d^2 v_{1s}}{d \xi^2} + \frac{1}{\xi} \frac{d v_{1s}}{d \xi}
+ \left( 1 - \frac{1}{\xi^2} \right) = 0,
\end{equation} 
where
\begin{equation}
\xi = ks \sqrt{F} \;\;, \;\;\; F = \frac{4 \omega_{A} \omega_{B}^2}{(\sigma^2
+ \omega_{A}^2)^2} -1.
\end{equation}
The solution of this equation can be represented in terms of the Bessel
functions of the order of 1 (see, e.g., Morse \& Feshbach 1953),
\begin{equation}
v_{1s} = C_1 J_{1}(\xi) + C_2 Y_{1}(\xi),
\end{equation}
with $C_1$ and $C_2$ being constant. For the chosen boundary conditions, 
the eigenvalues of Eq.~(13) are determined by the equation
\begin{equation}
J_{1}(\xi_1) Y_{1}(p \xi_1) - Y_{1}(\xi_1) J_{1}(p \xi_1)=0, 
\end{equation}
where $p = s_2/s_1$ and $\xi_1 = ks_1 \sqrt{F}$. The roots of Eq.~(16)
are real and simple. If $p >1$, the asymptotic expansion of the $n$th 
zero is
\begin{equation}
\xi_{1}^{(n)} \approx \frac{n \pi s_1}{\Delta s} + \frac{3 \Delta s}{8 n \pi 
s_2} + ... \;\;, \;\;\; \Delta s =s_2 - s_1 
\end{equation}
(see, e.g., Olver 1970). Since $\xi_{1}^{(n)} = ks_{1} \sqrt{F}$, we obtain 
the dispersion relation for $\sigma$ in the form
\begin{equation}
(\sigma^2 + \omega_{A}^2)^2 = 4 \mu \omega_{A}^2 \omega_{B}^2,  
\end{equation}
where $\mu = k^2/(k^2 + k_s^2)$ and $k_s= \xi_{1}^{(n)}/s_1$. 
Then,
\begin{equation}
\sigma^2 = -\omega_{A}^2 \pm 2 \sqrt{\mu} \omega_{A} \omega_{B}.
\end{equation}
One of the roots is positive if the toroidal field satisfies the
inequality
\begin{equation}
B_{\varphi}(s_1) > B_z \left( \frac{k s_1}{2 \sqrt{\mu}} \right),
\end{equation}
where $B_{\varphi}(s_1)$ is the strength of the toroidal field at the
inner boundary. If condition (20) is fulfilled, then  two 
aperiodic modes exist with
\begin{equation}
\sigma_{1,2} \approx \pm \mu^{1/4}\sqrt{2 \omega_{A} \omega_{B}} \propto 
\sqrt{B_z B_{\varphi}}.
\end{equation} 
One of these modes is always unstable, and another one is stable and decays 
exactly on the timescale on which the unstable mode grows. The other 
two modes that exist in the considered magnetic configuration are also 
stable but oscillatory. Under condition (22), their frequencies are given 
approximately by
\begin{equation}
\sigma_{3,4} \approx \pm i \mu^{1/4}\sqrt{2 \omega_{A} \omega_{B}}, 
\end{equation}
and are also $\propto \sqrt{B_z B_{\varphi}}$. To the best of our knowledge, 
the type of MHD wave with dispersion equations (21) and (22) has not been 
considered in the literature yet. These modes exist in fluid only if both the 
axial and toroidal magnetic field are non-vanishing since their frequency (or 
growth/decay timescale) is proportional to $\sqrt{B_z B_{\varphi}}$. Note 
that if the toroidal field is weak and condition (20) is not satisfied, the 
modes given by Eq.~(18) transform into ordinary Alfv\'en waves.

{
\subsection{Instability of short wavelength perturbations}

If the wavelength of perturbations is shorter than the characteristic
lengthscale of unperturbed quantities, then one can neglect the terms of the
order $1/s$ compared to $d/ds$ in Eq.~(11). In this case, Eq.~(11) can be 
transformed into
\begin{eqnarray}
\frac{d^2 v_{1s}}{d s^2}  + \frac{2 \delta \omega_{A}^2}{s (\sigma^2 + 
\omega_A^2)}  \frac{d v_{1s}}{ds} - k^2 \left[ 1 - 
\frac{4 \omega_{A}^2 \omega_{B}^2}{(\sigma^2 + \omega_{A}^2)^2} 
\right. 
\nonumber \\
\left. + \frac{2 \omega_{B}^2  ( 1 - \alpha )}{(\sigma^2 + \omega_A^2)} 
\right] v_{1s} = 0.
\end{eqnarray} 
In a short-wavelength approximation, the coefficients of this equation
can be treated as constant, and the solution can be taken in the form $F 
\propto \exp(-ik_{s} s)$ where $k_s$ is the wavevector in the radial 
direction. The dispersion relation corresponding to Eq.~(23) is
\begin{equation}
(\sigma^2 + \omega_A^2)(\sigma^2 + \omega_A^2 + 2 A) - 4 \mu 
\omega_A^2 \omega_B^2 = 0,
\end{equation}   
where
$$
A = (1 - \alpha)  \mu \omega_B^2
+ \frac{i k_s}{s q^2} \delta \omega_A^2 \;, \;\; \mu = \frac{k^2}{Q^2} \;,
\;\; Q^2 = k^2 + k_s^2 \;. 
$$
The solution of Eq.~(24) is
\begin{equation}
\sigma^2 = - \omega_A^2 - A \pm \sqrt{A^2 + 4 \mu \omega_A^2
\omega_B^2 }.
\end{equation}
If the azimuthal field is vanishing ($\omega_B=0$), then the roots are
\begin{equation}
\sigma_{1, 2}^2 = - \omega_A^2 \;, \;\;\; \sigma_{3, 4}^2 = - \omega_A^2
- \frac{2 i k_s}{s Q^2} \delta \omega_A^2 \;.
\end{equation} 
The first two modes are always stable. The modes 3 and 4 
are usually stable since the second term in the expression 
for $\sigma_{3, 4}$ is of the order of $1/ (s k_s) \ll 1$ compared to the
first term and should be neglected in a short-wavelength approximation. The 
only exception is the region in a neighbourhood of the neutral point (or 
line) where $B_z(s) \rightarrow 0$ and, hence, $\omega_A \rightarrow 0$. 
For example, if $B_z(s)$ goes to zero $\propto (s-s_0)$, then the first term 
in $\sigma_{3, 4}^2$ is $\propto (s-s_0)^2$, but the second one is $\propto 
(s-s_0)$. Therefore, the region always exists in the neighbourhood of $s_0$ 
where the second term dominates the first one. In this region, the roots 3 
and 4 are approximately given by
\begin{equation}
\sigma_{3, 4} \approx \pm \frac{1-i}{Q} \sqrt{\frac{\delta k_s}{s}} \;
\omega_A.
\end{equation}
One of the roots (27) corresponds to the unstable mode. Therefore, poloidal
magnetic configurations with the neutral point (line) are always unstable
in the neighbourhood of this point (line). This fact was first pointed out
by Tayler (1973).

However, the presence of the azimuthal field can drastically change  the
stability properties of the magnetic field even in a neighbourhood of the
neutral point. Near the neutral point, the azimuthal field is stronger than 
the axial one (if the former does not have zero at the same point $s_0$), 
then from Eq.~(25) we have with the accuracy in terms of the order of 
$\omega_A^2$
\begin{equation}
\sigma_{1, 2}^2 \approx 2 (\alpha - 1) \mu \omega_{B}^2 \;,  \;\;\;
\sigma_{3, 4}^2 \approx \omega_A^2 \; \frac{1 + \alpha}{1 - \alpha}.
\end{equation}
It turns out that the stability is determined by the properties of the 
azimuthal field rather than the presence of a neutral point in the magnetic 
configuration. If $\alpha > 1$, then one of the modes (1, 2) is unstable. 
On the contrary, one of the modes (3, 4) is unstable if $\alpha < 1$. Note 
that the instability of the modes 3 or 4 is less efficient in the 
neighbourhood of the neutral point than in the rest volume (see below) because 
$\omega_A$ is small there and, hence, the growth time of instability is 
long.   

Consider now the local stability of a region that is far off the neutral 
point. In this case, the term in $A$ proportional to $\delta$ can be 
neglected in Eq.~(25). Then, we have 
\begin{equation}
\sigma^{4} + a_{2} \sigma^{2} + a_{0}= 0, 
\end{equation}
where 
$$
a_{2} = 2 \omega^{2}_{A} + 2 \mu \omega^{2}_{B} (1 - \alpha ), 
\;\;
a_{0} = \omega_{A}^2 [ \omega^{2}_{A} - 2 \mu \omega^{2}_{B} (1 + 
\alpha ) ]. 
$$ 
Equation~(29) has positive (unstable) roots if either $a_2 < 0$ or $a_0 < 0$. 
These two conditions are equivalent to
\begin{equation}
\alpha> 1 + (k s_1)^2 \frac{B_z^2}{\mu B_{\varphi}^2}
\end{equation}   
and
\begin{equation}
\alpha > - 1 + (k s_1)^2 \frac{B_z^2}{2 \mu B_{\varphi}^2},
\end{equation}
respectively. Obviously, if condition (31) is fulfilled, then condition (30) 
is also satisfied. Therefore, the true criterion of instability is given by 
Eq.~(31). Depending on the parameters, the sufficient condition of instability
(31) can differ substantially from the necessary condition (1) that is
equivalent to $\alpha > -1$.   
The roots of Eq.~(29) are 
\begin{equation}
\sigma^2 \! = \! - \! \omega_{A}^2 - \mu \omega_{B}^2 (1 - \alpha )
\pm \sqrt{\mu^2 \omega_{B}^4 (1 - \alpha )^2
+ 4 \mu \omega_{A}^2 \omega_{B}^2}.
\end{equation}
In the case of a weak axial field ($\omega_B > \omega_A$ or $B_{\varphi} > 
(ks_1) B_z$), the behaviour of instability is rather different for $\alpha > 
1$ and $\alpha < 1$. If $\alpha > 1$, the unstable root of Eq.~(32) is 
equal to
\begin{equation}
\sigma \approx \omega_B \sqrt{2 \mu (\alpha -1)}.
\end{equation} 
In this case, the instability grows on the timescale of the order of the time
it takes an Alfv\'en wave in the field $B_{\varphi}$ to travel to the 
distance $s_1$. This regime recalls the instability of a purely toroidal
field that occurs only at $\alpha > 1$ and grows on the same timescale $\sim
1/\omega_B$ (Braithwaite \& Nordlund 2006). If $\alpha < 1$, then the root 
corresponding to instability is given by 
\begin{equation}
\sigma \approx \sqrt{\frac{1+\alpha}{1 - \alpha}} \omega_A.
\end{equation}
For such profiles of the toroidal field, the instability essentially grows  
slower. Its growth time is of the order of an inverse Alfv\'en frequency 
associated with the axial magnetic field, $B_z$. This instability is less
efficient in a relative neighbourhood of the neutral point where $\omega_A$
is small. Note that the value $\alpha \approx 1$ distinguishes between two 
regimes only in the limit of weak $B_z$. If $B_z$ is stronger, then the 
distinguishing value becomes higher. 

Note that, in the case $\alpha =1$, Eq.~(32) recovers the dispersion relation
(19). If the axial magnetic field is sufficiently strong, axisymmetric 
perturbations are stable for any profile $B_{\varphi}(s)$.
}

\section{Numerical results}

In this section we discuss the stability of the magnetic configurations
containing axial and azimuthal fields by numerically solving the eigenvalue 
problem.  In calculations, we assume  that the dependence of the azimuthal 
magnetic field on $s$ is given by
\begin{equation}
B_{\varphi} = B_{\varphi 0} \left( \frac{s}{s_1} \right)^{\alpha},
\end{equation} 
where $B_{\varphi 0}$ is the field strength at the inner boundary. 
To calculate the growth rate of the instability, it is  convenient to 
introduce dimensionless quantities
\begin{equation}
x \!= \! \frac{s}{s_1},\;\; q \!= \! k s_1,\; \Gamma \! = \!
\frac{\sigma}{\omega_{B0}}, \; \omega_{B0} \!= \! \frac{B_{\varphi 0}}{s_1
\sqrt{4 \pi \rho}}, \;
\varepsilon \! = \! \frac{B_z}{B_{\varphi 0}}.  
\end{equation}
Then, Eq.~(11) transforms into
\begin{eqnarray} \label{pert}
\frac{d}{dx} \left( \frac{d v_{1s}}{dx} + \frac{v_{1s}}{x} \right)- q^2 
\left\{ 1 - x^{2 (\alpha-1)} \left[ \frac{4 q^2 \varepsilon^2}{(\Gamma^2 + 
q^2 \varepsilon^2)^2} \right. \right.
\nonumber \\
\left. \left. - \frac{2 (1 - \alpha)}{\Gamma^2 + q^2 \varepsilon^2} \right] 
\right\} v_{1s} =  - \frac{2 \delta}{x} \frac{q^2 \varepsilon^2}{\Gamma^2 +
q^2 \varepsilon^2} \left( \frac{\partial v_{1s}}{\partial x} + 
\frac{v_{1s}}{x} \right).
\end{eqnarray}
{ The parameter $\varepsilon$ can depend on $s$ in this equation since 
$B_z$ is generally a function of $s$ in our model.}  
The inner and outer boundaries correspond to $x_1 =1$ and $x_2 =2$, 
respectively, where we have $v_{1s}(x_1)=v_{1s}(x_2) =0$. 
Equation~(36), together with the given  boundary conditions at the extrema, is 
a two-point boundary value problem that can be solved by using the 
``shooting'' method (Press {\it et al.} 1992). In particular in order to 
solve Eq.~(36), we used a fifth-order Runge-Kutta integrator embedded in a 
globally convergent Newton-Rawson iterator.  As the spectrum is discrete, we 
first found the lowest eigenvalue for the  analytical solution (19), and 
then we gradually changed $q$, $B_z/B_\varphi$ and $\alpha$ in order to 
explore the parameter space. In this way we checked that the eigenvalue 
was always the fundamental one,  as the corresponding eigenfunction had no 
zero except the one at the  boundaries. Higher radial wavenumbers corresponds 
to lower growth rates, as it is also apparent from Eqs.(20)-(21).

\begin{figure}
\includegraphics[width=9.0cm]{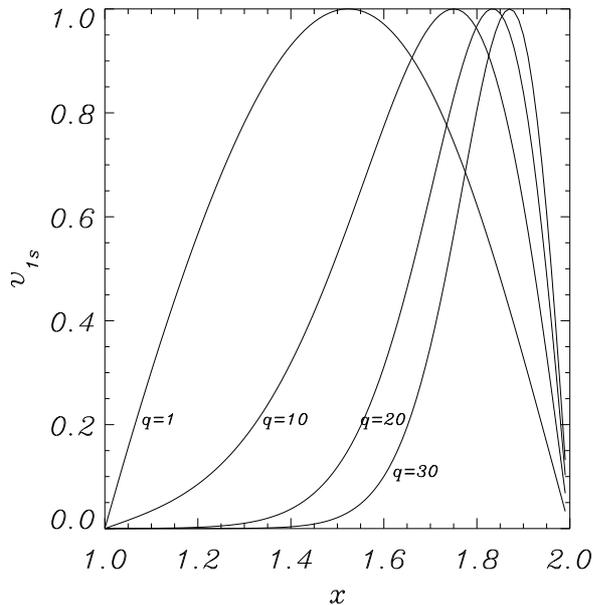}
\caption{The dependence of the fundamental eigenfunctions on $s$ for $\alpha
=2$, $B_z/B_{\varphi 0}=0.5$, and several values of $q$.}
\end{figure}

The radial dependence of the eigenfunctions is shown in Fig.~1 for $\alpha=2$
and $B_z/B_{\varphi} =0.5$ ($\delta =0$). The curve corresponding to $q=1$ 
describes the
unstable mode, but the other 3 curves represent stable modes. The
eigenfunctions for $q<1$, representing unstable modes as well, almost 
coincide with the curve $q=1$ and are not shown in Fig.~1. Note that, for 
other values of the parameters, the unstable eigenfunctions can also have 
sharp maximum near the outer (if $\alpha > 0$) or inner (if $\alpha < 0$) 
boundaries. It seems that the maxima tend usually to be located near the
boundary with the strongest field. 
  
\begin{figure}
\includegraphics[width=9.0cm]{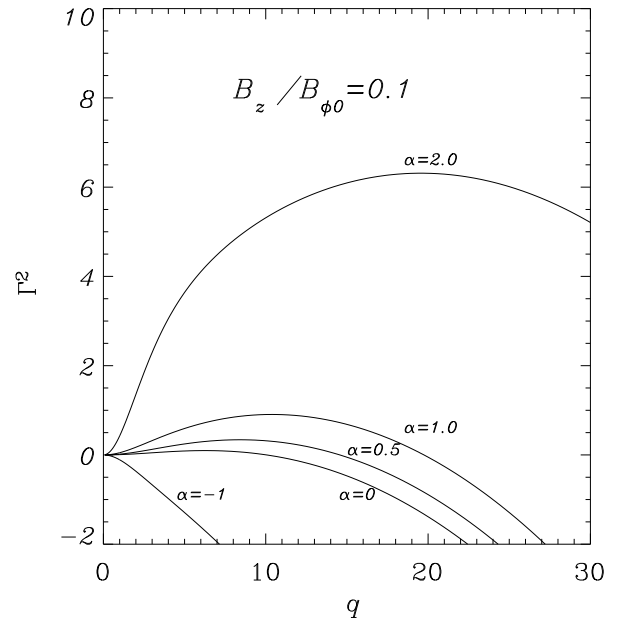}
\caption{The dependence of $\Gamma^2$ on $q$ for $B_z/B_{\varphi 0}= 0.1$
and several values of $\alpha$.}
\end{figure}
 
In Fig.~2, we plot $\Gamma=\sigma/\omega_{B0}$ as a function of $q=k s_1$ 
for a weak constant axial field. The addition of even very weak $B_z$ changes 
the stability properties substantially. A purely toroidal field should be
unstable if $\alpha > 1$. Our calculations show, however, that even the
profile of the toroidal field with $\alpha=0$ is unstable in this case 
despite a very low energy of the axial field compared to that of the 
toroidal field ($\sim 1$\%). It is also worth noticing  an important
difference between the necessary condition of instability (1) and the 
sufficient condition. In accordance with inequality (1), the necessary 
condition requires $\alpha > \alpha_{c} = - 1$, but our calculations show 
that the instability occurs at essentially larger $\alpha_{c} \sim -0.1$. The 
growth rate as a function of $q$ always has a smoothed maximum at $q \sim 
5-20$ depending on $\alpha$. The value of $q$ corresponding to the maximum 
increases with increasing $\alpha$. The figure clearly illustrates  the 
difference between the two regimes of the instability discussed in Sect. 4. 
If $\alpha < 1$, the instability is rather weak and grows on the Alfv\'en 
timescale characterised by the axial field. In contrast, if $\alpha 
> 1$, the instability is much more efficient, and the growth time is 
close to the Alfv\'en timescale for the toroidal field. 

\begin{figure}
\includegraphics[width=9.0cm]{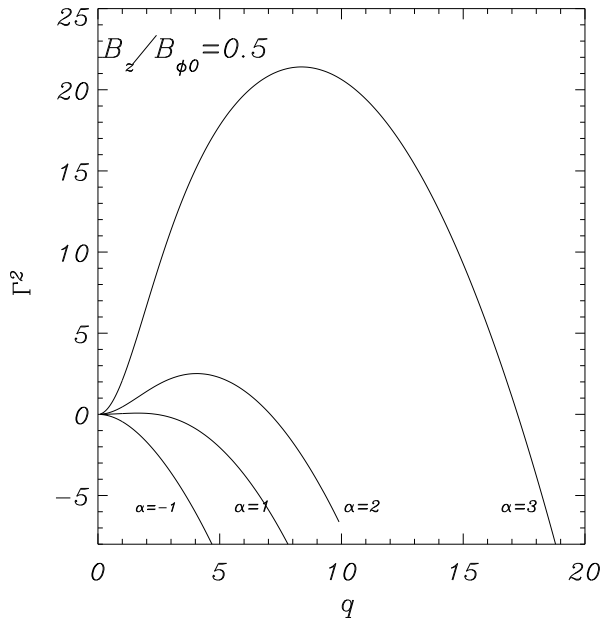}
\caption{The same as in Fig.~2 but for $B_z/B_{\varphi 0}=0.5$.}
\end{figure}

The dependence of $\Gamma^2$ on $q$ for a stronger axial field is shown
in Fig.~3. An increase of the axial field makes the magnetic configuration
more stable. For example, in the case $\alpha=2$, the maximum growth rate
of instability is $\approx 2.5 \; \omega_{B 0}$ if $\varepsilon=0.1$, but it
only becomes  $\approx 1.7 \; \omega_{B 0}$ if $\varepsilon =0.5$. As 
mentioned, the critical value of $\alpha$ that determines the onset of 
instability is $\sim -0.1$ in the case $\varepsilon =0.1$. If $\varepsilon =
0.5$, then instability can occur only if the toroidal field increases 
enough rapidly with $s$, and the critical value $\alpha_c$ is close to
$1$. In a stronger axial field, the transition between two regimes of
instability (see Eqs.~(30) and (31)) occurs at a higher  value of $\alpha$ as
was predicted in Section 3.2.  
 
\begin{figure}
\includegraphics[width=9.0cm]{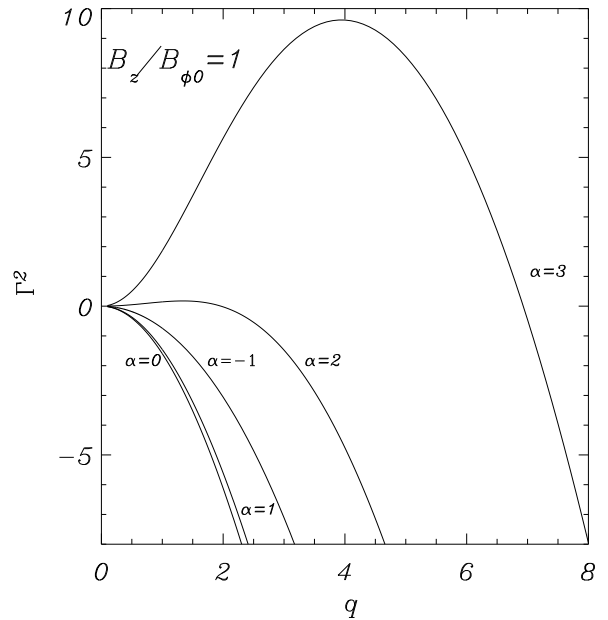}
\caption{The same as in Fig.~2 but for $B_z/B_{\varphi 0}=1$.}
\end{figure}

In Fig.~4, the normalized growth rate is shown for $B_z/B_{\varphi}=1$. Such
a strong axial field stabilises the magnetic configuration drastically. The 
instability occurs only if the toroidal field increases very rapidly with 
$s$ ($\alpha \geq 2$). A qualitative behaviour of unstable eigenvalues is the 
same as in Fig.~2 and 3, but the maximum growth rate corresponds to  lower
values of $q$.   

\begin{figure}
\includegraphics[width=9.0cm]{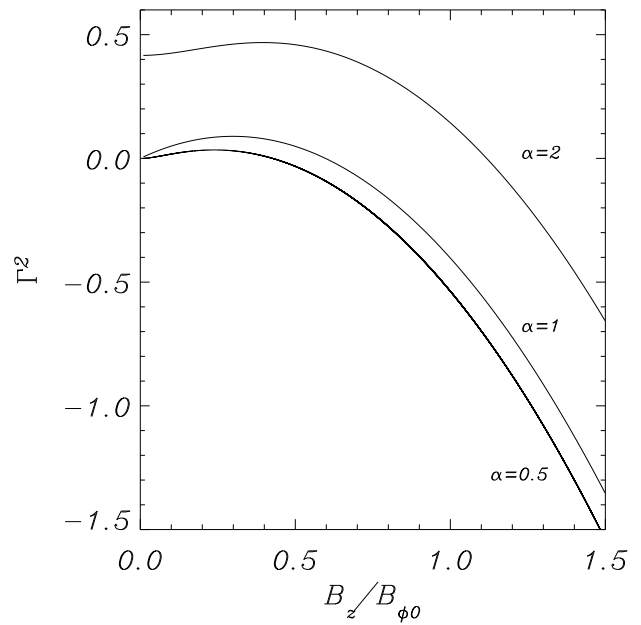}
\caption{The dependence of $\Gamma^2$ on $B_z/B_{\varphi 0}=1$ for $q=1$
and different values of $\alpha$.}
\end{figure}

In Fig.~5, we plot the dependence of $\Gamma^2$ on $\varepsilon = B_z/
B_{\varphi 0}$ for $q=1$. The profiles of the toroidal field with $\alpha
=0.5$ and $1$ are stable if $B_z=0$. Therefore, these curves start with
$\Gamma^2 =0$. In contrast, the toroidal field profile with $\alpha=2$
is unstable even if the axial field is vanishing and, therefore, $\Gamma^2
(\varepsilon = 0) \neq 0$. All the curves shown in Fig.~5 exhibit a 
qualitatively similar behaviour: the growth rate reaches a rather flat
maximum at $B_z/B_{\varphi 0} \approx 0.3-0.5$ and then decreases fast enough.
The quantity $\Gamma^2$ becomes negative and, hence, the magnetic
configuration is stable if $B_z$ is sufficiently strong. The higher is the 
value of $\alpha$, the stronger the axial field  required for stabilization. 

\begin{figure}
\includegraphics[width=9.0cm]{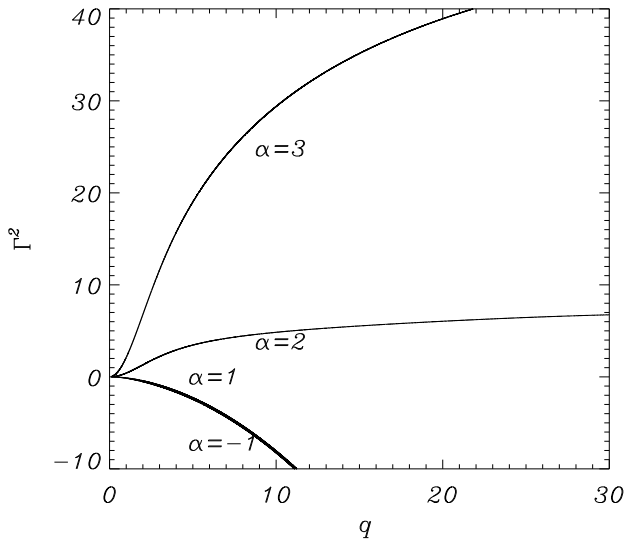}
\caption{The same as in Fig.~2 but for $B_z$ given by Eq.~(38) with $x_0=
1.75$ and $B_{z0}/B_{\varphi 0}=0.6$.}
\end{figure}

{ In Fig.~6, we plot the dependence of $\Gamma^2$ on $q$ for the axial
magnetic field that depends on $s$. We choose this dependence in the form
\begin{equation}
B_z(s) = B_{z0} \left( 1 - \frac{s}{s_0} \right) =
B_{z0} \left( 1 - \frac{x}{x_0} \right),
\end{equation}
where $x_0 = s_0/s_1$, $s_0$ is a radius at which $B_z$ changes  sign,
$s_2 > s_0 > s_1$. We have for such $B_z$
\begin{equation}
\delta = \frac{x}{x - x_0} \;, \;\;\; \varepsilon = \varepsilon_0
\frac{x_0 -x}{x_0} \;,
\end{equation} 
where $\varepsilon_0 = B_{z0}/B_{\varphi 0}$. Calculating models in Fig.~6,
we suppose $x_0 = 1.75$ and $\varepsilon_0 = 0.6$. The ratio $B_z /
B_{\varphi 0}$ averaged over $s$ is equal to 0.1 in this case, thus the 
results can be compared with those in Fig.~2 where $B_{z}=$const and $B_z/
B_{\varphi 0}=0.1$. Although the axial field has the neutral line at $s_0 = 
1.75$, the stability properties are not very different from the
similar configuration with no neutral line (Fig.~2). In Fig.~6, the critical 
value of $\alpha$ that distinguishes between stable and unstable 
configurations is close to $0$. If the axial field has no neutral line 
(Fig.~2), then the critical value is fairly close to this value ($\sim -0.1$). 
The maximum growth rate is of the same order of magnitude for the same 
$\alpha$. The only difference is that configurations with the neutral line
(Fig.~6) are unstable for a wider range of $q$. 
 
\begin{figure}
\includegraphics[width=9.0cm]{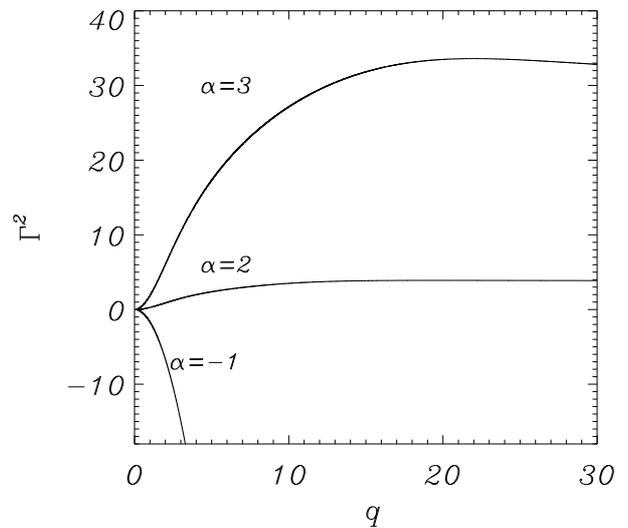}
\caption{The same as in Fig.~6 but for $B_{z0}/B_{\varphi 0}=3$.}
\end{figure}

In Fig.~7, we plot $\Gamma^2$ versus $q$ for the case of a stronger axial 
field ($B_{z0}/B_{\varphi 0}=3$) with the same neutral line at $s_0=1.75$.  
The qualititative behaviour is similar to what is depicted in Fig.~6, although
the growth rates are lower because of the higher value of $B_{z0}/
B_{\varphi 0}$. The averaged ratio $B_{z}/B_{\varphi 0}$ is 0.5 in this case, 
and the results can be compared to Fig.~3 where $B_z/B_{\varphi 0}=0$.
It appears that the presence of a neutral line does not change 
the stability properties substantially. For instance, the critical value of $\alpha$ is 
$\sim 1$ for  both configurations. The maximum growth rate at a given
$\alpha$ is of the same order of magnitude as well. However, the rage of 
unstable $q$ is wider for configurations with the neutral line.
}

\section{Discussion}

{ We have considered the hydromagnetic stability of cylindrical 
configurations containing the toroidal and axial magnetic fields. Dissipative
effects were neglected in our study. We treated a linear stability assuming 
that the behaviour of small perturbations is governed by equations of
incompressible hydrodynamics. This approximation is  justified if the 
magnetic field is subthermal and the Alfv\'en velocity is low compared to 
the sound speed.} The stability of the magnetic configurations is a key issue 
for understanding the properties of various astrophysical bodies, such as 
peculiar A and B stars, magnetic white dwarfs, neutron stars, etc. { 
Even though  various dynamo models predict that the toroidal field should be 
typically stronger than the poloidal one, the effect of a poloidal field on 
the stability usually cannot be neglected.} 

To demonstrate this, we treated the simplest model of a highly conducting fluid
between two cylindrical surfaces. We assumed that the toroidal and axial 
fields depends on the cylindrical radius alone. In a short-wavelength 
approximation, we  derived the growth rate and a sufficient criterion of 
instability analytically (Eqs.~(30)-(31)). For large-scale 
perturbations, the condition of instability and its growth rate were 
calculated numerically. The analytical and numerical results are in  good 
qualitative agreement. The obtained conditions of instability differ 
substantially from what is predicted by the necessary condition (1). For 
instance, according to Eq.~(1), if the instability occurs in the magnetic 
configuration, then the toroidal field profile satisfies the condition 
$\alpha > -1$. In fact, the instability occurs only if the toroidal field 
decreases with $s$ much slower (or even increases): the critical value of 
$\alpha$ is $\approx -0.1$ if $B_z/ B_{\varphi 0} =0.1$ and $\approx 1$ if 
$B_z/ B_{\varphi 0} =0.5$. { If $B_{z}$ depends on $s$, then the critical
values of $\alpha$ should be even higher.} 
 
Depending on the profile of the toroidal field and the strength of the 
axial field, the instability can arise in two essentially different
regimes. In the case of a weak axial field, $B_{\varphi 0} \gg B_z$, the 
value of $\alpha$ that distinguishes between the regimes is $\approx 1$. If 
$\alpha > 1$, then the instability grows on the Alfv\'en timescale determined 
by the toroidal field and is rather fast. If $\alpha < 1$, then the 
instability is much slower and grows on the timescale determined by the
axial field. The transition between these two regimes occurs at larger 
$\alpha$ if the axial field increases. The efficiency of the considered 
instability turns out to be rather low if $\alpha < 1$ and $B_z$ is weak.   

It is worth noticing the very particular properties of instability in the case 
$\alpha \approx 1$. In such a configuration, the particular type of MHD 
waves is given by the dispersion equation (19). The growth rate of these 
waves (or the frequency, if the waves are stable) is proportional to the 
product of $B_z$ and $B_{\varphi}$. These waves cannot exist in purely 
toroidal or purely poloidal fields. The instability of the configuration 
with $\alpha \approx 1$ is caused by the generation of this particular type 
of wave. { These waves can probably determine the instability of magnetic
configurations near the axis of symmetry where $B_{\varphi} \rightarrow 0$.}  

A sufficiently strong axial field always suppresses the instability. For 
more or less plausible values of $\alpha \leq 1$, the strength of the 
axial field stabilising the configuration is $\sim 0.1-1 B_{\varphi}$. A 
much stronger field is required, however, to stabilise the
configuration with larger $\alpha$.

\vspace{0.5cm}

\noindent
{\it Acknowledgments.}
This research project was  supported by a Marie Curie Transfer of
Knowledge Fellowship of the European Community's Sixth Framework
Programme under contract number MTKD-CT-002995.
VU thanks also INAF-Ossevatorio Astrofisico di Catania for hospitality.

{}

\end{document}